\documentclass[iop]{emulateapj}
\usepackage{graphicx}
\usepackage{amsmath}

\newcommand{\ud}{\mathrm{d}}

\newcommand{\Teff}{T_{\rm eff}}
\newcommand{\NH}{N_{\rm H}}
\newcommand{\Mwd}{M_{\rm WD}}
\newcommand{\Lbol}{L_{\rm bol}}
\newcommand{\Lgeff}{\log(g_{\rm eff})}

\newcommand{\phx}{{\texttt {PHOENIX}}}

\begin{document}

\title{A Public Set of Synthetic Spectra from Expanding Atmospheres for X-Ray Novae. I. Solar Abundances}
\author{Daniel R. van Rossum}
\affil{Department of Astronomy and Astrophysics, University of Chicago, Chicago,IL 60637, USA}
\affil{Center for Astrophysical Thermonuclear Flashes, University of Chicago, Chicago,IL 60637, USA}
%\altaffiltext{2}{Center for Astrophysical Thermonuclear Flashes, University of Chicago, Chicago, IL 60637}


\begin{abstract}
X-ray grating observations have revealed great detail in the spectra of Novae in the Super Soft Source (SSS) phase.
Notable features in the SSS spectra are blue-shifted absorption lines, P-Cygni line profiles, and the absence of strong ionization edges, all of which are indicators of an expanding atmosphere.

We present, and make publicly available, a set of 672 wind-type synthetic spectra, obtained from the expanding NLTE SSS models introduced in \cite{Vanrossum10} with the \phx{} stellar atmosphere code.
The set presented in this paper is limited to solar abundances with the aim to focus on the basic model parameters and their effect on the spectra, providing the basis upon which abundance effects can be studied using a much bigger non-solar set in the next paper in this series.

We fit the wind-type spectra to the five grating spectra taken in the SSS phase of nova V4743 Sgr 2003 as an example application of the wind-type models.
Within the limits of solar abundances we demonstrate that the following parameters are constrained by the data (in order of decreasing accuracy): column density $\NH$, bolometric luminosity $\Lbol$, effective temperature $\Teff$, white dwarf radius $R$, wind asymptotic velocity $v_\infty$, and the mass-loss rate $\dot{M}$.
The models are also sensitive to the assumed white dwarf mass $\Mwd$ but the effect on the spectra can largely be compensated by the other model parameters.
The wind-type spectra with solar abundances fit the data better than abundance optimized hydrostatic models.
\end{abstract}

%-----------------------------------------------------------------------
\section{Introduction}
Novae are cataclysmic variables caused by a white dwarfs in a binary systems that experience a thermonuclear runaway in a compact surface-layer of hydrogen that was accreted from the companion.
The outburst is detected as a strong sudden increase in optical brightness followed by a gradual decline over a time scale of months.
Nova light curves vary substantially from system to system \citep{Strope10}.
Some novae go through a Super Soft Source (SSS) phase, in which spectra are measured similar to those of Super Soft X-ray binary Sources \citep{Kahabka97}.

Since the advent of the X-ray grating observatories, Chandra and XMM-Newton, `high'-resolution X-ray spectra have been obtained for a number of novae in their SSS phase.
These spectra contain a wealth of detail which have provided evidence of continued mass loss during the SSS phase: blue-shifted absorption lines, the absence of sharp ionization edges, and P-Cygni line profiles \citep{Ness07}.

These data so far were interpreted with the help of hydrostatic atmosphere models \citep{Rauch10, Osborne11, Ness11}.
These models have proven to be a powerful tool for determining effective temperatures and chemical compositions for grating observations.
But in order to match observed blue-shifted absorption lines the synthetic spectra are blue-shifted in an ad-hoc fashion \citep{Nelson08, Rauch10, Ness11}, which is not physically consistent with having a velocity field in a radially extended (vs. a compact hydrostatic) atmosphere.
The effects of the velocity field on the physical conditions of the atmosphere (e.g. density and temperature stratification) can significantly change the spectrum and affect the interpretation of the observations as shown by \cite{Vanrossum10}.

\cite{Petz05} introduced expanding atmosphere models for the SSS phase calculated with the \phx{} code \citep{Hauschildt92, Hauschildt99, Hauschildt04}.
Those models are similar to the \phx{} models used to simulate early UV and optical nova spectra \citep{Hauschildt92} but applied to the very high temperature conditions of SSS.
Although these models naturally reproduce blue-shifted absorption lines, neither the lines nor the continuum of the synthetic spectra match the shape of the observations well.

In \cite{Vanrossum10} a wind-type (WT) expanding atmosphere model was introduced for X-ray novae using the same \phx{} code.
Those models were shown to provide reasonable fits to two different SSS novae (RS Oph and V2491 Cyg), but their models were called preliminary and have since then neither been applied and examined, nor been available for the public to use for that purpose.

In this paper a large set of WT spectra is presented (and made publicly available), their properties explored, and their use demonstrated by application to the five V4743 Sgr 2003 grating observations.
V4743 Sgr 2003 was thoroughly\footnote{The abundances were tuned to individual observations.} analyzed with hydrostatic atmospheres in \cite{Rauch10}, so that the results presented here can be directly compared to previous work.

%-----------------------------------------------------------------------
\section{The ``wind-type'' (WT) SSS Nova model atmosphere setup}
The WT model atmospheres are based on two observations.
Firstly, (at least) a part of the atmosphere is expanding.
Secondly, given the high temperatures and small effective radii typically determined for SSS spectra (e.g. \cite{Osborne11}, \cite{Ness11}, \cite{Vanrossum10}), (at least) a part of the atmosphere is very close to the compact surface of the white dwarf.
Based on these two observations the WT model atmospheres assume a spherically symmetric compound structure with a hydrostatic base and an expanding envelope.
The hydrostatic base is described with the classical parameters effective temperature $T_{\rm eff}$, gravitational acceleration $\log(g)$, and radius $R$, using the hydrostatic equilibrium condition
\begin{equation}
 \frac{\ud p}{\ud r} = -\rho g  \qquad {\rm for} \quad r\le R\;,
\end{equation}
where $r$ is the radial distance to the center of mass, $R$ the outer radius of the static core, $p$ is the gas pressure, and $\rho$ the density of the gas.
The envelope is parameterized by a constant mass-loss rate $\dot{M}$ and a beta-law velocity field (see e.g. \cite{Lamers97})
\begin{equation}
 v(r>R) = v_0 + (v_\infty - v_0) \left(1 - \frac{R}{r}\right) ^\beta \;, \label{eq:EnvelopeV}
\end{equation}
with $v_\infty$ the asymptotic velocity, and $v_0$ the velocity at the base of the wind.
The parameter $\beta$ determines the length-scale in which the terminal velocity is reached.
The density of the envelope follows from the continuity equation for a constant mass loss rate
\begin{equation}
 \rho(r>R) = \frac{\dot{M}}{4\pi r^2 v(r)} \;. \label{eq:EnvelopeRho}
\end{equation}
The base velocity $v_0$ is refined in the model to provide a smooth run of the density and its gradient over the transition boundary at $r = R$ (see \cite{Thesis09}).
This is not possible for any arbitrary value of $\beta$.
Experiment shows that $\beta = 1.2 \pm 0.05$ for the WT models to have a smooth transition between static core and expanding envelope\footnote{
Note that the model atmosphere, and thus it's spectrum, depends on the assumed velocity law.  The effect of $\beta$ on the spectra is small within the allowed range.  The effect of assuming a different type of velocity-law is potentially bigger.  This is not explored within the scope of this work.}.
This is slightly higher than the value of $\beta = 0.9 \pm 0.1$ for radiation driven winds of blue giants (see e.g. \cite{Pauldrach86}).

In conclusion, the model atmosphere is defined by five parameters\footnotemark, the three classical hydrostatic parameters: $T_{\rm eff}$, $\log(g)$, $R$, and two parameters for the expanding envelope: $\dot{M}$, $v_\infty$.
\footnotetext{The chemical composition adds a large number of additional free parameters to the model, the discussion of which is deferred to next paper of this series.}

%------------------------------------------------------------------------
\section{Characteristics of the WT model} \label{sec:ModelCharacteristics}
With five parameters, the WT models have much more freedom than static atmospheres (which have three).
The effect of each of these parameters on the model atmosphere was examined in \cite{Thesis09}.
Here we demonstrate the effect on the observables, i.e. the spectra, focusing on four abstract properties:
\begin{enumerate} \setlength{\parskip}{0pt}
 \item Luminosity (or brightness)
 \item Color (or blueness)
 \item Blue-shift of absorption features
 \item Emission wing prominence in P Cyg profiles
\end{enumerate}
In the following, the details of the relations between model parameters and these observables is given, and an overview is shown in the Table \ref{tab:ParameterEffects}.
This table serves as a guide in the process of fitting WT spectra to data.
\begin{table}
\begin{center}
\caption{Overview on the effects of the five model parameters on four observables.} \label{tab:ParameterEffects}
\begin{tabular}{c|ccccc}
 & $R$ & $T_{\rm eff}$ & $\log(g)$ & $v_\infty$ & $\dot{M}/v_\infty$ \\
 \hline
 Luminosity & $\Uparrow$ & $\uparrow$ & -- & -- & -- \\
 Color & -- & $\Uparrow$ & $\downarrow$ & -- & $\uparrow$  \\
 Blue-shift & -- & -- & -- & $\Uparrow$ & $\uparrow$ \\
 Emission wings& -- & -- & -- & -- & $\Uparrow$ \\
\end{tabular}
\end{center}
{\bf Notes:}
{\footnotesize This table serves as a guide in the process of finding good spectral fits to data.
 $\uparrow$ means a positive correlation, $\downarrow$ a negative correlation, and -- no correlation between the model parameter and the observable (abstract property of the spectrum).
 The double arrows $\Uparrow$ denote the parameters that are most characteristic for each observable.\\
 The parameter $\log(g)$ has no double arrows meaning that it is not well constrained by the observables.
 The Color is affected by three parameters, which means that $\Teff$ generally is not determined uniquely, but depends on the value for $\log(g)$.}
\end{table}

\paragraph{Radius}
The atmosphere is self-similar with variation of $R$, meaning that it does not affect the flux produced by the model.
Yet, the total luminosity
\begin{equation}
 \Lbol = 4\pi R^2 \sigma_{\rm sb} T_{\rm eff}^4 \;,
\end{equation}
with $\sigma_{\rm sb}$ the Stefan-Boltzmann constant, scales trivially with $R^2$.

\paragraph{Effective Temperature}
$\Teff$ makes the spectrum both brighter and bluer (or dimmer and redder) in a way similar to how a blackbody spectrum changes with increasing (or decreasing) temperature.
But, in addition, the ionization balance of all chemical species changes with temperature which has a strong influence on the spectra.
The reason is that with effective temperatures typical for SSS models the C, N and O atoms are almost completely ionized to bare nuclei, which do not contribute to the opacity.
If the ionization balance changes slightly from bare nuclei towards H-like C, N and O then bound-free transitions start to significantly contribute to blue opacities making the spectrum redder.

\begin{figure}
 \centerline{\includegraphics[width=.48\textwidth]{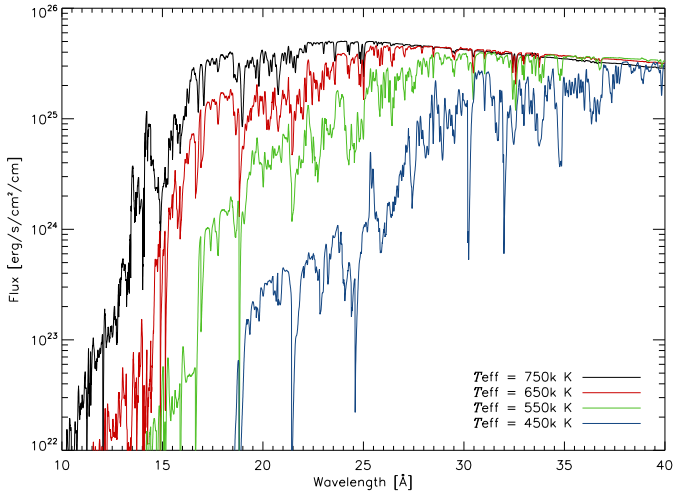}}
 \caption{The effect of $\Teff$ on the color of the WT model spectra.
 The fluxes are scaled to identical luminosities with $(750{\rm k K}/\Teff)^{4}$.\\
 The color of the spectrum changes bluer with increased $\Teff$ much more quickly than a blackbody does.
 This is caused by the high temperatures ionizing key elements like carbon, nitrogen and oxygen to bare nuclei, removing their bound-free opacities from the blue side of the spectrum if the temperature is increased even further.} \label{fig:Teff} 
\end{figure}

The brightness variation with effective temperature can completely be compensated by the radius $R$, so that $\Teff$ essentially is a color parameter.
The color variation of spectra with $\Teff$ is shown in Figure \ref{fig:Teff}, where spectra are scaled with $\Teff^{-4}$ to identical luminosities.

\paragraph{Surface Gravity}
Lowering $\log(g)$ makes the atmosphere less compact, so the density gradient becomes smaller.
As mentioned above, the wind base-velocity $v_0$ is refined in the model to provide a smooth run of the density and its gradient over the core-envelope boundary.
A smaller density gradient in the base of the envelope is obtained with higher $v_0$ values (from Equations \eqref{eq:EnvelopeV} and \eqref{eq:EnvelopeRho}), so that the transition density becomes lower.
Consequently, a lower surface gravity gets a wind that is optically thinner, the hot compact core moves ``up'' on the optical depth scale, and the spectrum becomes bluer.
The reverse holds for increasing $\log(g)$.

\begin{figure}
 \centerline{\includegraphics[width=.48\textwidth]{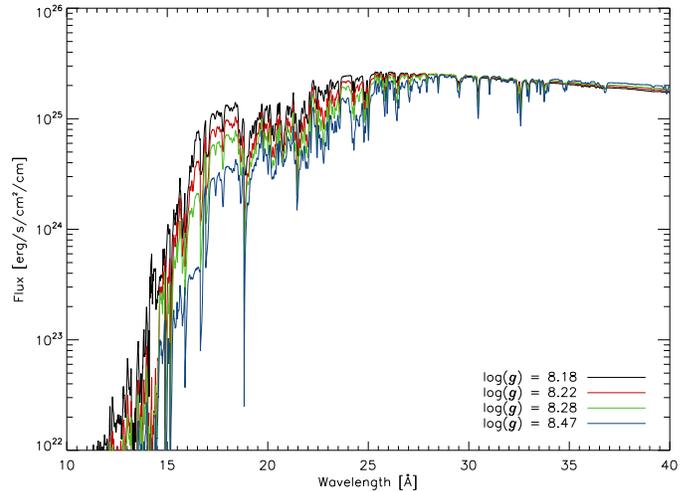}}
 \caption{The effect of $\log(g)$ on the WT spectra.
 A lower surface gravity in the model gets a matching wind that is optically thinner, as explained in the text.
 Consequently, the compact, hot, static part of the atmosphere moves ``up'' on the optical depth scale leading to a bluer color of the spectrum.\\
 Although this is a significant effect per se, the associated range in underlying white dwarf masses is very large (a factor of 2.2).
 Furthermore, the color effect of the surface gravity can largely be compensated by the color effect of the effective temperature as shown in Figure \ref{fig:LoggTeff}.
 This leaves $\log(g)$ poorly constrained by the observables.} \label{fig:Logg} 
\end{figure}

\begin{figure}
 \centerline{\includegraphics[width=.48\textwidth]{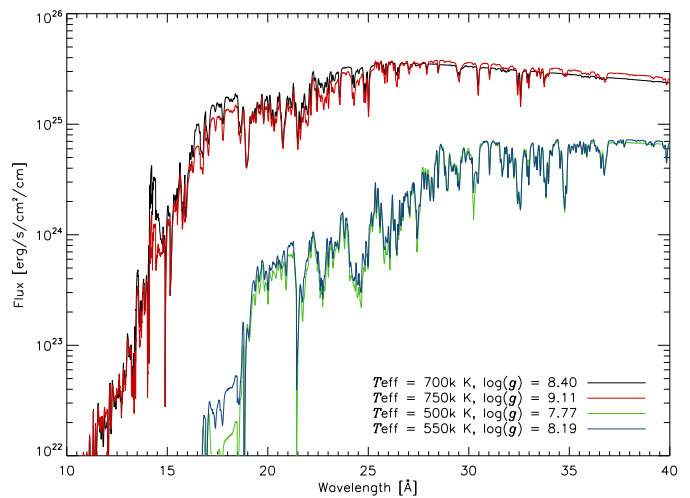}}
 \caption{The color effect of the surface gravity on the WT spectra (see Figure \ref{fig:Logg}), can largely be compensated by the color effect of the effective temperature (see Figure \ref{fig:Teff}).
 The fluxes of each pair of synthetic spectra are scaled to identical luminosities with $(700{\rm k K}/\Teff)^{4}$ (black-red) and  $(500{\rm k K}/\Teff)^{4}$ (green-blue) respectively, which corresponds to changing $R$.\\
 The underlying white-dwarf mass differs by a factor of 5.1 (black-red) and 2.7 (green-blue) respectively.
 $\log(g)$, and therefore $\Mwd$, are poorly constrained by the observables.} \label{fig:LoggTeff} 
\end{figure}

The assumed white dwarf mass follows from the values for $\log(g)$ and $R$ as
\begin{equation} \label{eq:Mwd}
 \Mwd = \frac{g R^2}{G}
\end{equation}
where $G$ is the gravitation constant.
In Figure \ref{fig:Logg} the color variation of spectra with $\log(g)$ is shown.
Given the large spread in white dwarf masses the influence is rather small.
Furthermore, as shown in Figure \ref{fig:LoggTeff}, the color effect of the surface gravity can largely be compensated by the color effect of the effective temperature, leaving $\log(g)$ poorly constrained by the observables.

\begin{figure}
 \centerline{\includegraphics[width=.48\textwidth]{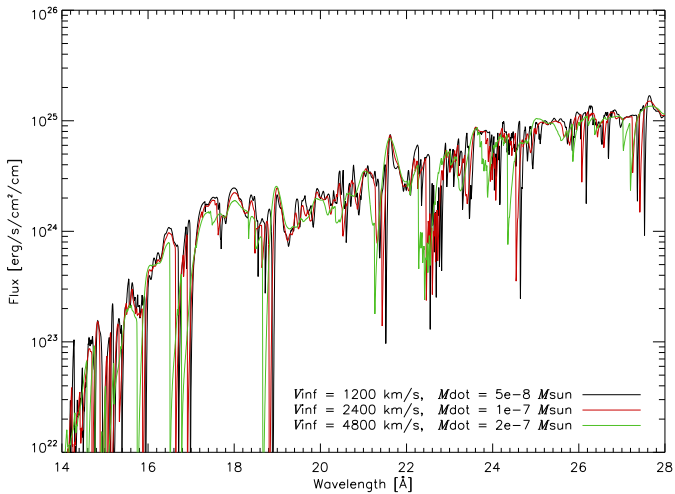}}
 \caption{The effect of $v_\infty$ on the WT spectra.
 In order to leave the density profile unaffected in the comparison $\dot{M}$ is scaled linearly with $v_\infty$, as described in the text.\\
 The line profiles become broader and the absorption part more blue-shifted with increasing $v_\infty$.
 Also, narrow features get more smeared out so that high-velocity spectra are smoother and low-velocity spectra show more small line features.
} \label{fig:Vinf}
\end{figure}

\paragraph{Asymptotic Velocity}
$v_\infty$ linearly scales the velocity law, apart from a small offset $v_0 \ll v_\infty$.
For a fixed mass-loss rate $\dot{M}$ the density of the wind changes with $v_\infty$.
But the density profile of the wind has a much bigger influence on the spectrum than the velocity profile.
In order to compare different velocities with identical density profiles the mass-loss rate is to be changed along with the terminal velocity, keeping $\dot{M}/v_\infty$ fixed.
In Figure \ref{fig:Vinf} the effect on the blue-shift of absorption profiles is shown.

\paragraph{Mass-loss Rate}
$\dot{M}$ is proportional to the wind density.
The emission wings of P Cyg profiles originate from the radially extended atmosphere, from those parts that are not in the line of sight to the compact core.
With increasing mass-loss rate the extended atmosphere becomes denser, therefore more material is available to line-scatter photons toward the observer, and so \emph{emission wings} of P Cyg profiles become stronger.

With more prominent emission wings, the absorption wing of the line profiles may appear to \emph{blue-shift} more strongly with increasing $\dot{M}$.

Another result of the increased wind density is that the transition to the compact core happens at higher densities and higher temperatures.
The temperature and density gradients in the wind are lower than in the static core, so that with $\dot{M}$ these increase throughout the whole wind region.
Both a higher temperature and a higher electron density boost ionization, shifting the balance towards bare nuclei and removing their opacity.
Consequently, the spectra turn \emph{bluer in color} as the mass-loss rate increases.

\begin{figure*}
 \centerline{\includegraphics[width=\textwidth]{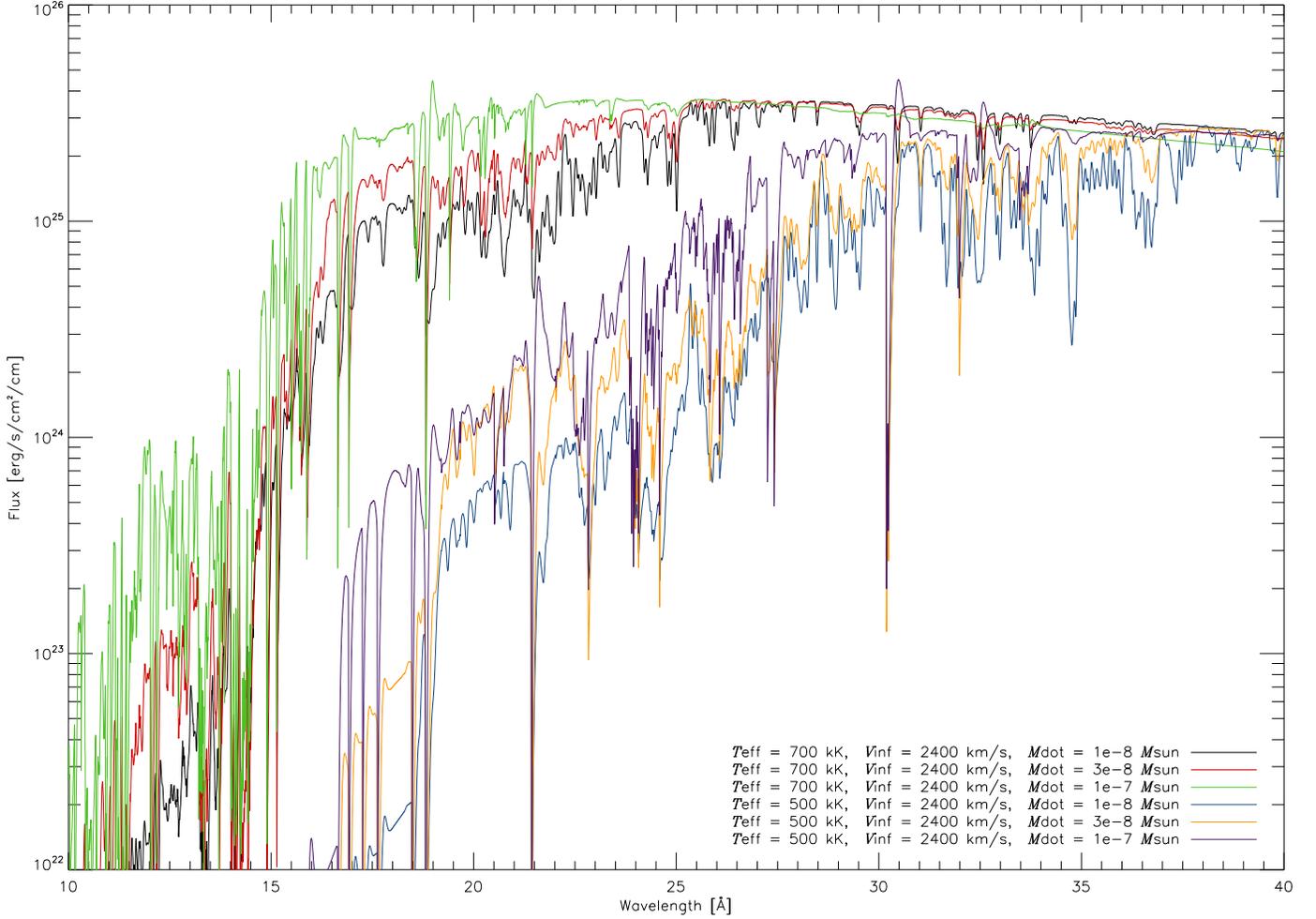}}
 \caption{The effect of $\dot{M}$ on the WT spectra.
 As described in the text, three of the abstract observables are affected by the mass-loss rate.
 Firstly, the emission wings of P Cyg profiles get stronger.
 Secondly, the blue-shift of absorption features increases slightly.
 Thirdly, the color of spectra turns bluer.\\
 Since the spectral color and the blue-shift can be partly compensated by $\Teff$ and $v_\infty$ respectively, the most characteristic effect of $\dot{M}$ is the prominence of emission wings in P Cyg profiles.\\
 The fluxes are scaled to identical luminosities with a factor of $(700{\rm k K}/\Teff)^{4}$.
 } \label{fig:Mdot}
\end{figure*}

All three effects $\dot{M}$ has on the spectra are shown in figure \ref{fig:Mdot}.
The spectral color and the blue-shift can be partly compensated by $\Teff$ and $v_\infty$ respectively.
Therefore, the most characteristic effect of $\dot{M}$ is the prominence of emission wings in P Cyg profiles.

%-----------------------------------------------------------------------
\section{Comparison with observations} \label{sec:ComparisonMethods}
\subsection{Interstellar absorption: \emph{NOT} a free parameter}
The colors of SSS spectral observations are strongly affected by absorption from the interstellar material (ISM).
In the comparison with observed data the synthetic flux is to be multiplied with the wavelength dependent transmission coefficient.
Generally, this coefficient depends on the ISM column depth $\NH$ and the chemical composition of the ISM.
In this paper we use the ISM absorption cross-section routines from \cite{Balucinska92} with solar (or almost-solar\footnote{
For the V4743 Sgr 2003 data, carbon and nitrogen abundances of [C] = -1 and [N] = 0.4 are assumed in order to match the non-absorbed fluxes on the right-hand and left-hand side of either ionization edge. All other abundances are assumed to be solar.}%
) abundances.

The ISM-absorbed spectrum is very sensitive to the $\NH$ parameter, especially the slope of the red tail (the part redder than about the peak wavelength) of the spectrum.
Although the red tail also is affected by the WT model parameters $\Teff$, $\log(g)$, and $\dot{M}$, all of those parameters have a bigger impact on the blue half of the spectrum.
We find, as a consequence, that the value of $\NH$ can accurately be determined, separately from the WT model parameters.
However, this can only be done if the red tail is completely measured by the instrument, like with the LETGS spectrograph on board CHANDRA.
The RGS spectrograph on board XMM-Newton is limited to 38.2\AA{} and therefore does not capture the red tail of the SSS spectrum, leaving $\NH$ undetermined.

Fit procedures that treat $\NH$ as a free parameter are sensitive to the wavelength range, as shown e.g. in \cite{Rauch10}, Figure 12.
That same Figure 12 also shows that chi-squared methods do not accurately determine $\NH$ even when the red tail is available, as evident from the non-matching slopes for wavelengths greater than 30\AA{}.
The problem is that the red tail has very little counts so that chi-squared methods, which optimize the fit in the high-counts bins, are not sensitive to an accurate slope in the red tail of the spectrum.
For this reason we claim that \emph{$\NH$ should not be treated like a free parameter}.
Doing so gives the fit too much freedom and leads to incorrect fit parameters.
Instead, once $\NH$ is determined from the long wavelength tail (provided this is possible for the data), fitting the short wavelength range with $\NH$ fixed will give more realistic fit parameters.

\begin{figure*}
 \centerline{\includegraphics[width=\textwidth]{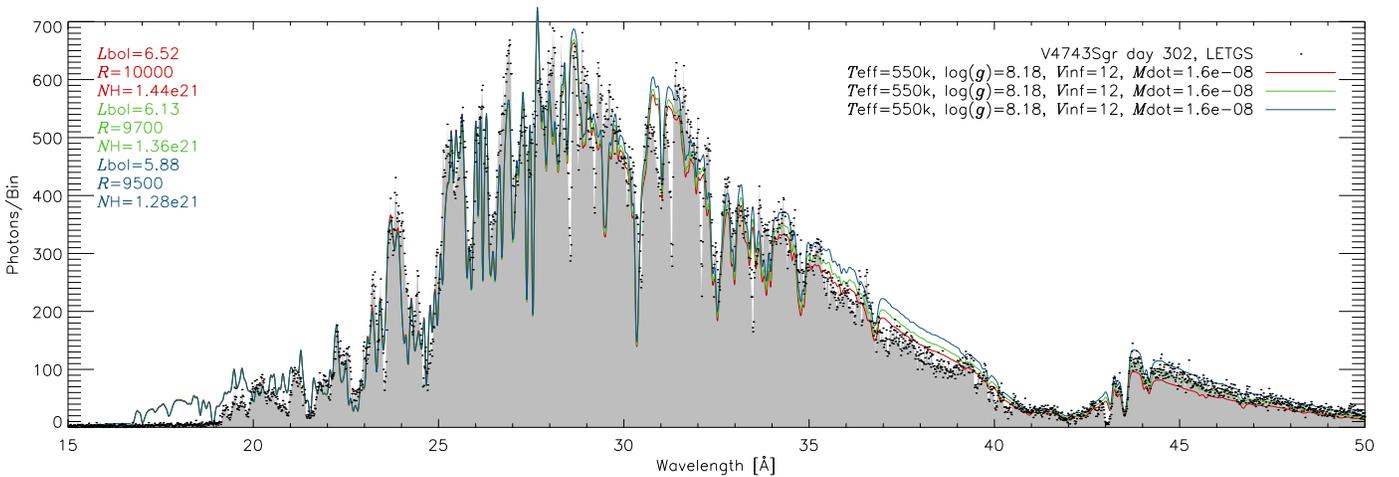}}
 \caption{The ISM absorption parameter $\NH$ strongly affects the comparison of synthetic spectra with data.
 The plot shows a V4743 Sgr (day 302) count spectrum compared with three WT spectra using different values of the interstellar absorption $\NH$: 1.44 (red), 1.36 (green), and 1.28 (blue) in units of $10^{21} {\rm cm}^{-2}$.
 The values of $\Teff$, $v_\infty$, $\dot{M}$, $\Lbol$, and $R$ shown in the plot are in the units: K, km/s, $M_\odot$/yr, erg/s, and km.\\
 $\NH$ determines the slope of the red tail of the spectrum.
 In this case the observed red tail slope is reproduced with $\NH = 1.36 \pm 0.04 \times 10^{21} {\rm cm}^{-2}$ (green curve).
 } \label{fig:v4743_nh}
\end{figure*}

The sensitivity of the ISM-absorbed spectrum to $\NH$ is demonstrated in Figure \ref{fig:v4743_nh}, where a synthetic spectrum is compared with an observation using three different values.
In this example $\NH$ can be determined with an uncertainty of $\pm 3\%$.
The associated uncertainties in $R$ and $\Lbol$ are $\pm 1.5\%$ and $\pm 2.5\%$ respectively.

\subsection{Counts spectra, Flux spectra, or both}
Comparisons between observations and synthetic spectra are often presented in the ``as-observed'' representation, as ISM-absorbed counts per bin on a linear scale, or in the ``as-modeled'' representation, as ISM-absorption corrected flux on a logarithmic scale.
Both representations have their advantages.

In the ``as-observed'' representation synthetic spectra are 'forwardly' corrected for ISM-absorption and convolved with the instrumental response of the detector.
The counts per bin numbers indicate how well each bin was exposed, preserving the statistical nature of X-ray measurements.
Note that any chi-squared fits in this representation rely on the \emph{arbitrariness} of the detector properties and the ISM absorption.

In the ``as-modeled'' representation observations are 'reversely' corrected for ISM-absorption and the instrumental response of the detector.
Deconvolution of the instrumental response is complicated and is compromised by small-number statistics.
Given that the observation is exposed well enough this representation shows the true flux.
Sure enough, a good model needs to not only reproduce the arbitrary (instrumental response and ISM absorption) high-count bins, but the whole run of the true flux from the source.
Also, the logarithmic scale better highlights the dim blue end of the spectrum which is sensitive to details of the atmosphere (e.g. temperature, mass-loss and composition).

For these reasons, we advocate to show comparisons in both the ``as-observed'' and the ``as-modeled'' representations.

%------------------------------------------------------------------------
\section{The WT model parameter grid} \label{sec:Grid}
With this paper a moderately sized parameter grid of WT model spectra is introduced and made publicly available\footnote{http://flash.uchicago.edu/\~{}daan/WT\_SSS/}.
It is a solid four-dimensional parameter grid covering the four WT model parameters other than $R$ (since variation with $R$ is trivial as explained in Section \ref{sec:ModelCharacteristics}).
For improved stability of the WT model $\log(g)$ is not directly specified but instead of $g$ the effective gravitational acceleration $g_{\rm eff} = g - a_{\rm rad}$ (at the surface of the static core), where $a_{\rm rad}$ is the acceleration from radiation pressure.
Therefore, the parameter grid is spaced in terms of $\Lgeff$.
Because the radiation pressure roughly scales with $\Teff^4$ we also scale $g_{\rm eff}$ with $\Teff^4$.
This way a gravitationally loosely bound atmosphere remains loosely bound along the $\Teff$ grid axis, i.e. the ratio $g/a_{\rm rad}$ is roughly independent of $\Teff$.

The parameter grid size is $8 \times 7 \times 4 \times 3$ in the dimensions $\Teff$, $\Lgeff$, $v_\infty$, and $\dot{M}$ respectively, for a total of 672 spectra.
The parameter values at the grid points are listed in Table \ref{tab:ModelGridPoints}.
\begin{table}
\begin{center}
\caption{WT spectra are calculated on these parameter grid points.} \label{tab:ModelGridPoints}
\begin{tabular}{c|c|c}
 Param. &\#& Grid point values\\
 \hline
 $\Teff$ &8& [450, 475, 500, 550, 600, 650, 700, 750] $\times 10^3$ K\\
 $\Lgeff$ &7& -15.55 + $\Teff^4$ + [0.0, 0.2, 0.4, 0.6, 0.8, 1.0, 1.2] \\
 $v_\infty$ &4& [1200, 2400, 3600, 4800] km/s\\
 $\dot{M}$ &3& $v_\infty/2400 \,\times$ [$10^{-8}$, $3 \cdot 10^{-8}$, $10^{-7}$] $M_\odot$/yr
\end{tabular}
\end{center}
\end{table}

%------------------------------------------------------------------------
\section{Uncertainties from $\Mwd$-ambiguity of WT models} \label{sec:MassAmbiguity}
\begin{figure*}
 \centerline{\includegraphics[width=\textwidth]{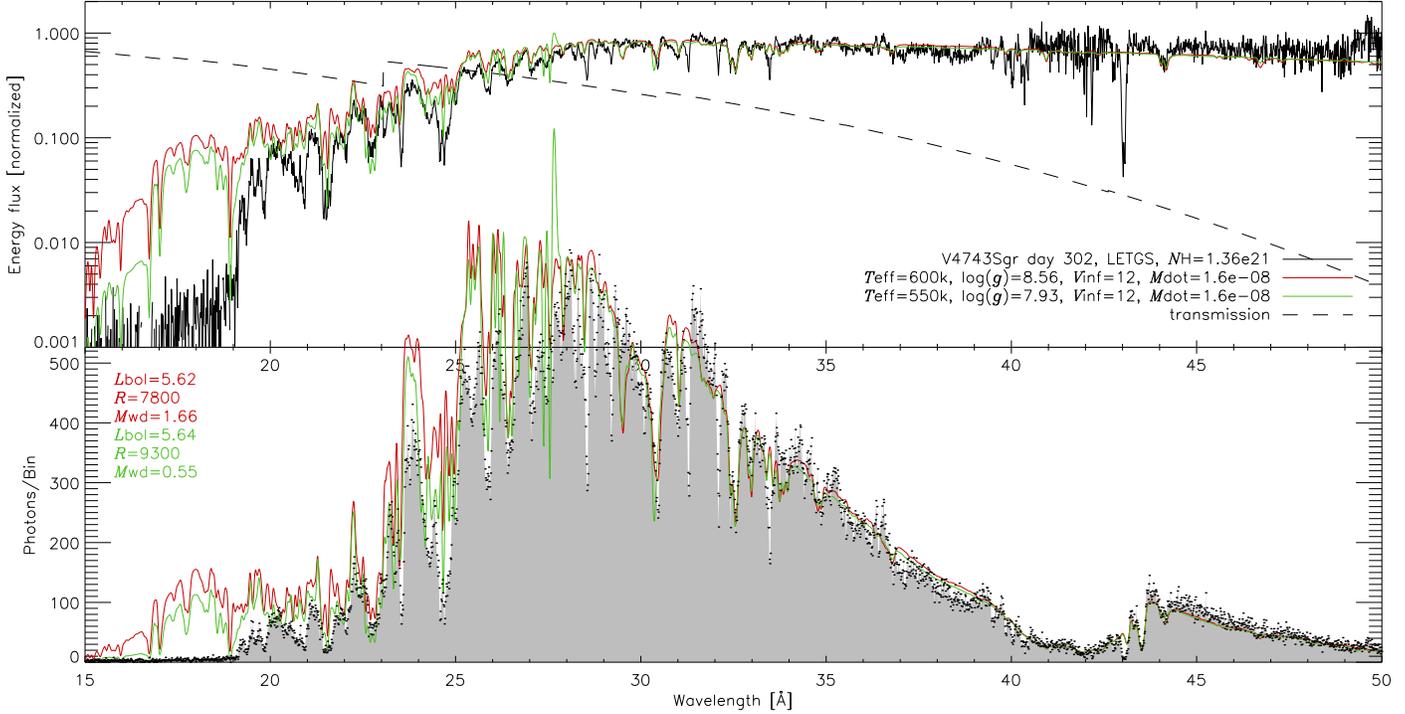}}
 \caption{The fit of two WT spectra with different $\log(g)$ values shows that observed data can be equally well fit with very different white dwarf masses of 0.47$M_\odot$ (green curve) and a 1.42$M_\odot$ (red curve).
 The comparison with data from V4743 Sgr (day 203) is shown in two representations, a logarithmic unabsorbed flux plot called ``as-modeled'' (upper-panel) and a linear ISM-absorbed counts plot called ``as-observed'' (lower panel).
 The values of $\Teff$, $v_\infty$, $\dot{M}$, $\Lbol$, and $R$ shown in the plot are in the units: K, km/s, $M_\odot$/yr, erg/s, and km.\\
 The $\Mwd$-ambiguity causes uncertainties in the fit parameters $\Teff$ and $R$, while the determined value of $\Lbol$ is little affected.
 } \label{fig:v4743_mass}
\end{figure*}

In Section \ref{sec:ModelCharacteristics} it was explained that $\log(g)$ affects the WT spectra (Figure \ref{fig:Logg}), but that the effect can be largely compensated with $\Teff$ (Figure \ref{fig:LoggTeff}).
But with $\Teff$ the luminosity $\Lbol$ of the model changes which can be compensated by changing $R$.
Thus, the white dwarf mass changes with $\log(g)$ and $R$ simultaneously (see Equation \eqref{eq:Mwd}).
The ambiguity of the white dwarf mass in the WT models is demonstrated in Figure \ref{fig:v4743_mass}, where a 0.47$M_\odot$ and a 1.42$M_\odot$ model give an equally good description of the data.
The differences between the WT spectra are too small to favor one above the other, especially considering the enormous difference in white dwarf mass.

The $\Mwd$-ambiguity imposes uncertainties in the correlated quantities $\Teff$ and $R$, the size of which is to be determined from fits to the data.
From Figure \ref{fig:v4743_mass} we determine that the uncertainty from $\Mwd$-ambiguity is $\pm 5\%$ in $\Teff$ and about $\pm 8\%$ in $R$.
The uncertainty from $\Mwd$-ambiguity in the bolometric luminosity $\Lbol$ is approximately $\pm 1.5\%$.

The assumption of solar abundances causes additional systematic uncertainties, the sizes of which have to be determined from fits of non-solar models to the data, which is beyond the scope of this work.

%------------------------------------------------------------------------
\section{Fits to V4743 Sgr} \label{sec:Results}
\begin{figure}
 \centerline{\includegraphics[width=.48\textwidth]{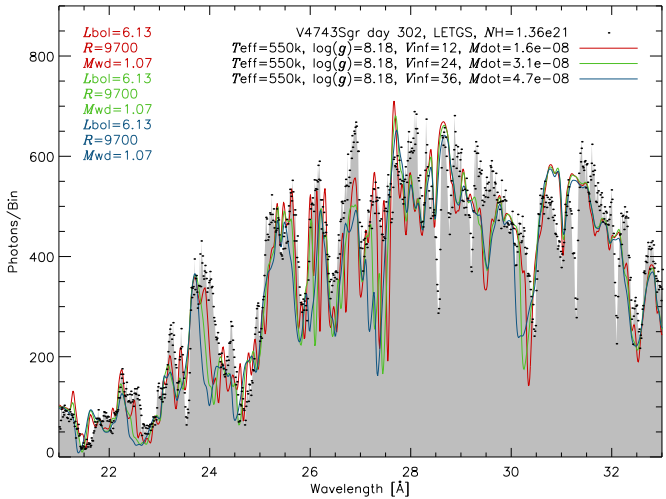}}
 \caption{Three WT model spectra with different wind asymptotic velocities compared to the V4743 Sgr Day 302 observation: $v_\infty$ = 1200 km/s (red), 2400 km/s (green), and 3600 km/s (blue).
 Note that the mass-loss rate changes along with $v_\infty$, as explained in section \ref{sec:ModelCharacteristics} and Figure \ref{fig:Vinf}.
 The units of the quantities shown are described in the caption of Figure \ref{fig:v4743_mass}.\\
 Some absorption features become increasingly blue-shifted with $v_\infty$, e.g those around 24.2, 26.3, and 27.5 \AA{}, whereas other remain at their rest wavelengths, e.g. those around 25.8, 28.2, and 31.1 \AA{}.
 Furthermore, the spectral features become increasingly washed-out, i.e. broader and shallower, with increasing $v_\infty$.\\
 The steeper spectral features and the moderate blue-shifts in the $v_\infty$=1200 km/s model give the best match to the data.
 } \label{fig:v4743_vinf}
\end{figure}

\begin{figure}
 \centerline{\includegraphics[width=.48\textwidth]{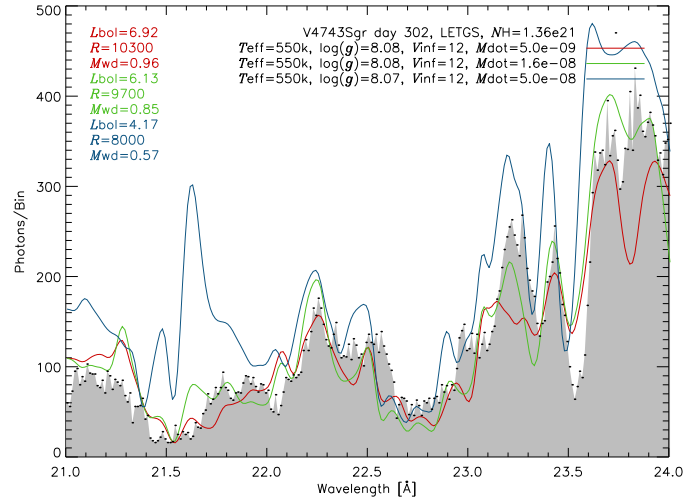}}
 \caption{Three WT model spectra with different mass-loss rates compared to the V4743 Sgr Day 302 observation: $\dot{M} = 5.0 \cdot 10^{-9} \, M_\odot/{\rm yr}$ (red), $1.6 \cdot 10^{-8} \, M_\odot/{\rm yr}$ (green), and $5.0 \cdot 10^{-9} \, M_\odot/{\rm yr}$ (blue).
 The units of the quantities shown are described in the caption of Figure \ref{fig:v4743_mass}.
 Note that the wind densities are relatively high given the low wind velocities, $v_\infty$ = 1200 km/s.\\
 With increasing mass-loss rate each isolated absorption feature gets a more prominent emission wing so that in the interplay of many overlapping lines the spectrum shows deeper and narrower features.
 In comparison with the data, this effect is too strong for the highest mass-loss rate model and too weak for the lowest.\\
 Within the available parameter grid spacing a value of $\dot{M} = 1.6 \cdot 10^{-8} M_\odot / {\rm yr}$ reproduces the observations best.
 The match in detail also depends on the chemical abundances which are not optimized in the scope of this work.
 } \label{fig:v4743_Mdot}
\end{figure}

Figure \ref{fig:v4743_vinf} and \ref{fig:v4743_Mdot} demonstrate the determination of $v_\infty$ and $\dot{M}$ from the Day 302 spectrum of V4743 Sgr.

With increasing wind velocity some absorption features get increasingly blue-shifted while other lines only change in profile shape, depending on where in the atmosphere the lines are formed.
Lines that become optically thick in the outer wind layers are more blue-shifted than those formed deeper inside.
In the high-velocity models the blue-shift in the lines that are affected is stronger than what is observed for V4743.
Also, in those models the line profiles are too washed-out (or smeared-out), which is most apparent in the lines that do not blue-shift.

If the mass-loss rate increases each absorption line gets a more prominent emission wing.
The combined effect of many such P Cygni profiles leads to an increasing number of peaks and pits in the spectra, and steep slopes between them.
At the same time, the spectra become bluer (see section \ref{sec:ModelCharacteristics}), which is compensated by lowering the effective temperature.
In turn the white dwarf radius is increased to maintain the overall flux level, and the surface gravity changes in order to compare models with approximately equal white dwarf masses.
The relative uncertainty in $\dot{M}$ is 50\% and the associated additional uncertainties in $\Teff$ and $R$ are 5\%.

\begin{figure*}
 \centerline{\includegraphics[width=\textwidth]{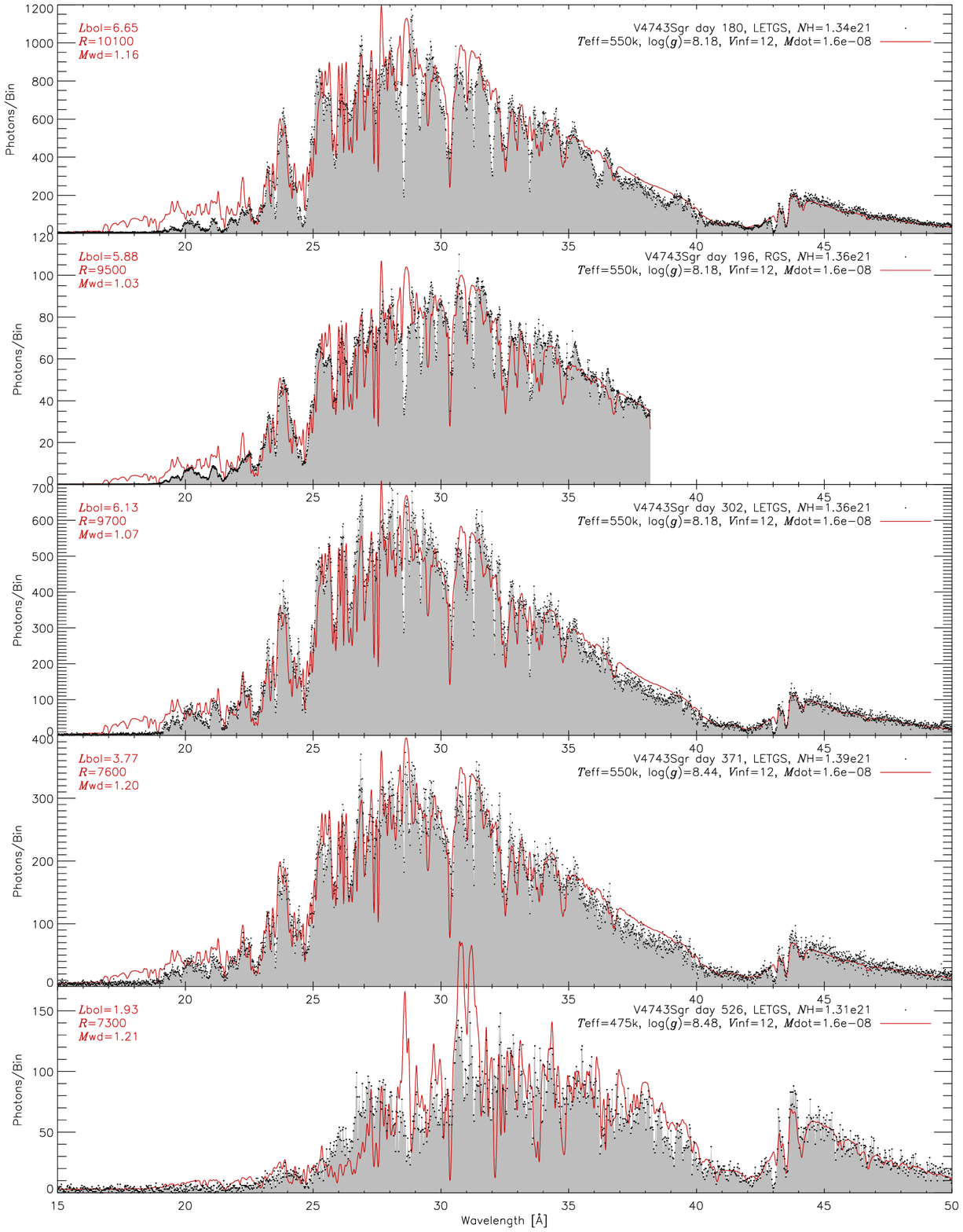}}
 \caption{WT model spectra fit to V4743 Sgr grating data in the ``as-observed'' representation (see section \ref{sec:ComparisonMethods}).
 The values of $\Teff$, $v_\infty$, $\dot{M}$, $\Lbol$, and $R$ shown in the plot are in the units: K, km/s, $M_\odot$/yr, erg/s, and km.\\
 } \label{fig:v4743_cts}
\end{figure*}

\begin{figure*}
 \centerline{\includegraphics[width=\textwidth]{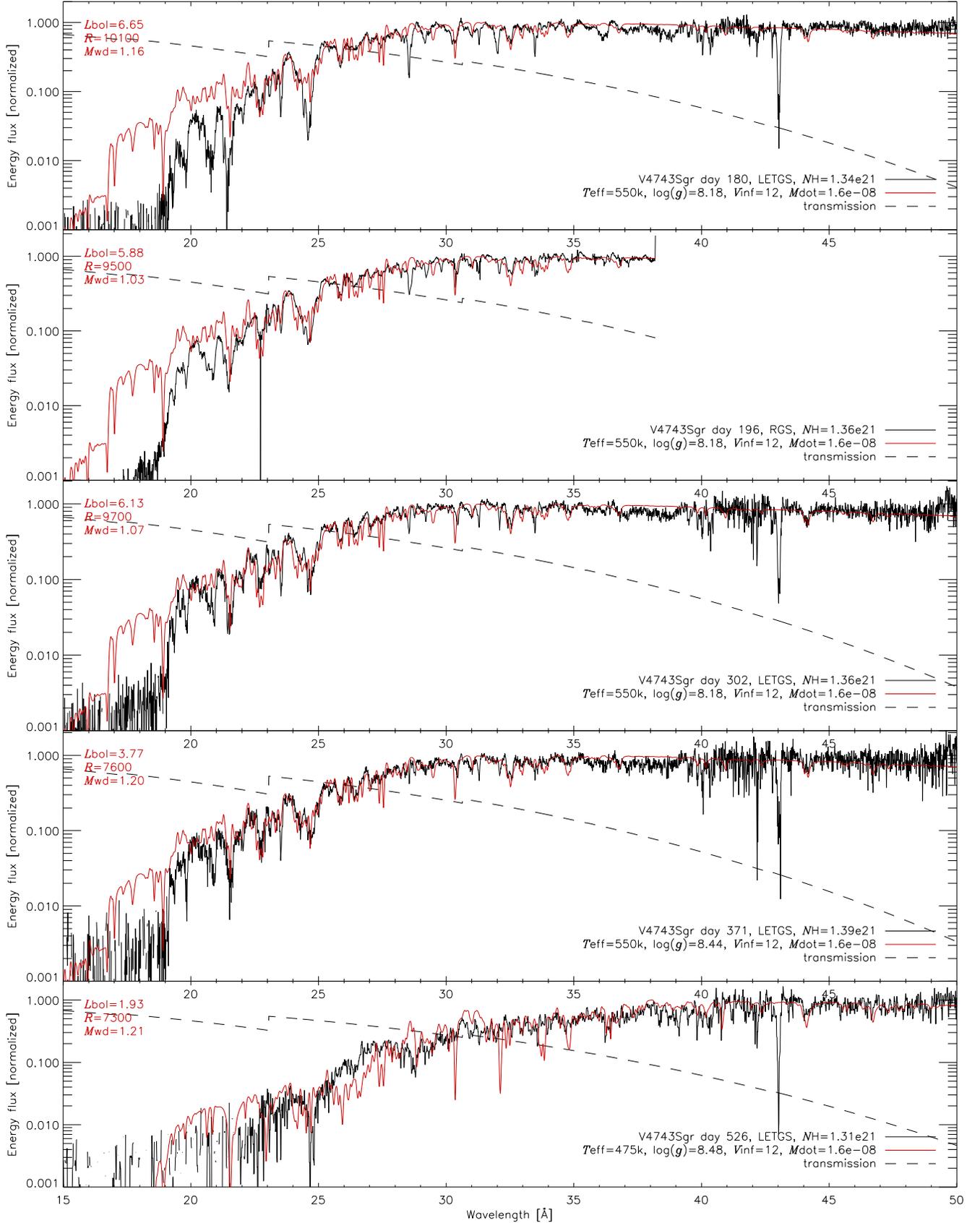}}
 \caption{WT model spectra fit to V4743 Sgr grating data in the ``as-modeled'' representation (see section \ref{sec:ComparisonMethods}).
 The units of the quantities shown are described in the caption of Figure \ref{fig:v4743_cts}.} \label{fig:v4743_flx}
\end{figure*}

\begin{figure*}
 \centerline{\includegraphics[width=\textwidth]{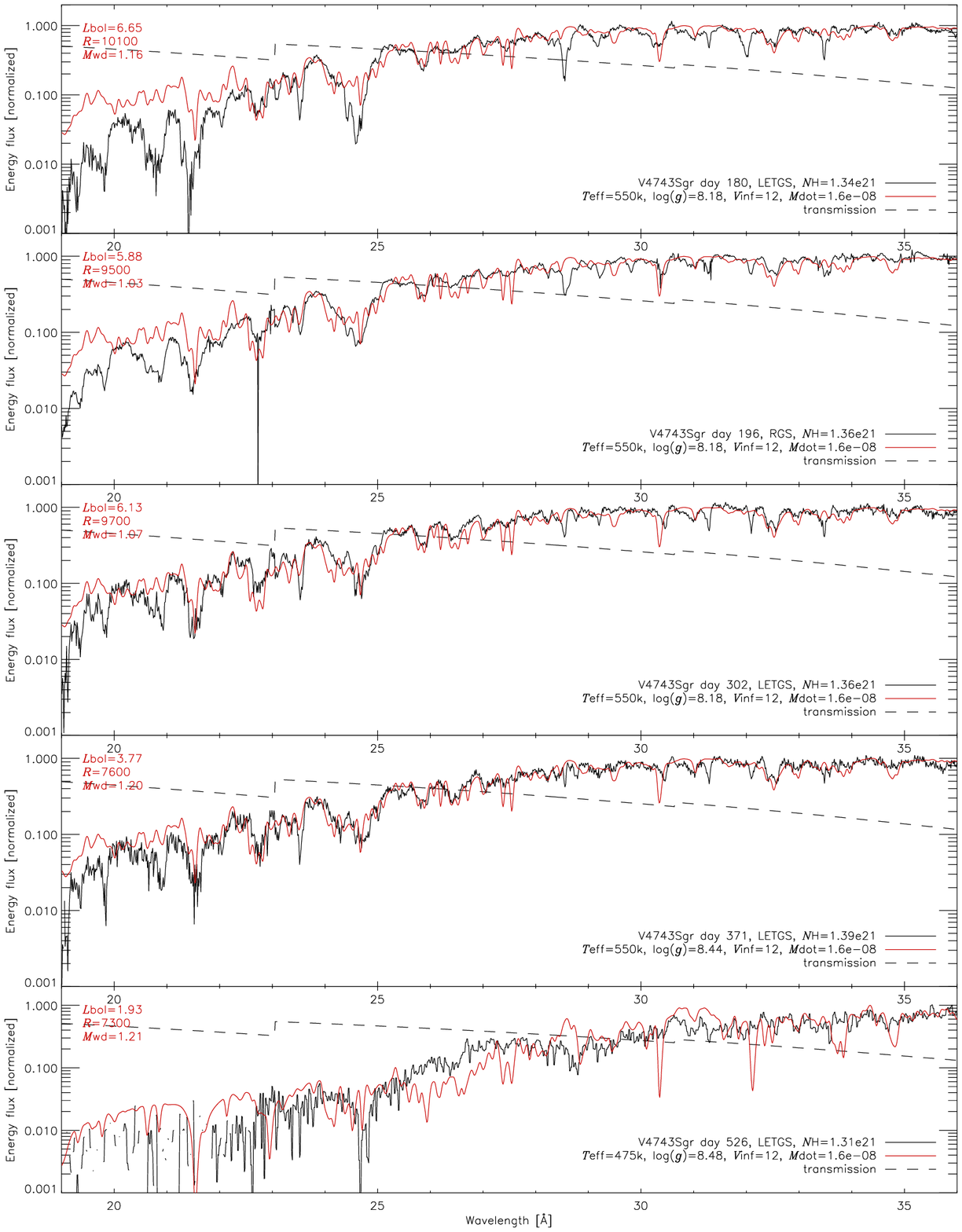}}
 \caption{Like the previous figure, but with a smaller wavelength range, zooming in on the detailed line profiles in the spectra.} \label{fig:v4743_flx_zoom}
\end{figure*}

The fit results for all five grating spectra available for V4743 Sgr are shown in Figures \ref{fig:v4743_cts} to \ref{fig:v4743_flx_zoom}.
The white dwarf masses (for which the WT models are degenerate, see section \ref{sec:MassAmbiguity}) in each of the fits are chosen around an arbitrary value of $\Mwd = 1.1 M_\odot$ as close as possible within the available grid spacing (see section \ref{sec:Grid}).

The Day 526 observation is less well exposed than the first four observations.
We re-binned the Day 526 data by a factor of two to suppress the noise.
The Day 196 observation was made with the RGS spectrograph on board XMM-Newton with an upper wavelength limit of 38.2\AA{}.
For this observation no individual determination of $\NH$ is possible (as explained in section \ref{sec:ComparisonMethods}).
Therefore, we assumed the average value from the other observations.

The best-fit parameters and relative uncertainties are listed in Table \ref{tab:FitParameters}.
\begin{table}
\begin{center}
\caption{Best fit parameters and uncertainties for all five grating observations of V4743 Sgr.} \label{tab:FitParameters}
\begin{tabular}{c|cccccc}
 %Day & $\NH$ & $\Lbol\,^a$ & $\Teff \,^b$ & $R\,^c$ & $v_\infty \,^d$ & $\log(\dot{M}) \,^e$ \\
 Day & $\NH$ & $\Lbol$ & $\Teff$ & $R$ & $v_\infty$ & $\dot{M}$ \\
 \hline
 180 & 1.34 & 6.65 & 5.5 & 10. & 1.2 & 1.6 \\
 196 &   -- & 5.88 & 5.5 & 9.5 & 1.2 & 1.6 \\
 302 & 1.36 & 6.13 & 5.5 & 9.7 & 1.2 & 1.6 \\
 371 & 1.39 & 3.77 & 5.5 & 7.6 & 1.2 & 1.6 \\
 526 & 1.31 & 1.93 & 4.8 & 7.3 & 1.2 & 1.6 \\
 \hline
 %units & $10^{21}$ cm$^{-2}$ & $10^{37}$ erg/s & $10^5$ K & $10^8$ cm & $10^8$ cm/s & $M_\odot$/yr \\
 uncert. & $\pm$3\% & $\pm$3\% & $\pm$8\% & $\pm$11\% & $\pm$30\% & $\pm$50\%
\end{tabular}
\end{center}
{\bf Notes:}\\
{\footnotesize
 The uncertainties do not include the effects from abundance variations, which are not explored in the scope of this work, nor the distance uncertainty.
 The assumed distance to V4743 Sgr is 4kpc (see text).\\
 $\NH$ is in units of $10^{21}$ cm$^{-2}$.  ISM abundances are assumed to be solar, except $[C]$=-1.0 and $[O]$=0.4 (see text). \\
 $\Lbol$ is in units of $10^{37}$ erg/s \\
 $\Teff$ is in units of $10^5$ K \\
 $R$ is in units of $10^8$ cm \\
 $v_\infty$ is in units of $10^8$ cm/s \\
 $\dot{M}$ is in units of $10^{-8} M_\odot$/yr\\
}
\end{table}
The distance to V4743 is not accurately known in literature.
\cite{Nielbock03} determine an upper limit of 6kpc from optical decline rate considerations and a range of 0.4-1.2kpc using reference stars, but also point out that these values are uncertain.
We assume a value of $D = 4$kpc as a ``golden mean'' guess.
Note, that values for $R$ and $\Lbol$ scale linearly and quadratically with $D$.
The uncertainties for $R$ and $\Lbol$ as listed in Table \ref{tab:FitParameters} only reflect effects that can be directly determined with the models.
For final values, the distance uncertainty has to be propagated accordingly.

%-----------------------------------------------------------------------
\section{Discussion and Conclusions}
The representation of the data by the models is good, especially for the first four observations.
Many, though not all, of the observed spectral features are (at least) roughly reproduced.
The overall flux level matches well over the whole wavelength range except shortward of the 22.5 \AA{} where the N {\sc VI} ionization edge is not strong enough in the models.
Shortward of 18.6 \AA{}, the N{\sc VII} ionization edge, the discrepancy grows even stronger.
Also, the strong N{\sc VI} and N{\sc VII} lines at 28.9 and 24.8 \AA{} are too weak in the models.
Obviously, a super-solar N abundance would improve these aspects of the fits to the data.
Note that the relative strengths of ionization edges and absorption lines are largely determined by NLTE effects.
The dependence of these on abundance variations is generally hard to predict.

The ISM column depth varies as little as 6\% over the four LETGS observations.

The first four observations are best-fit by a single effective temperature.
This would \emph{not} have been the case if the $\log(g)$ parameter would not have been adapted to satisfy a fixed underlying white dwarf radius.
The variations in bolometric luminosity can here be explained by only varying the white dwarf radius and adapting the surface gravity correspondingly.

The mass-loss rate and the asymptotic velocity of the wind do not significantly vary over the five grating observations of V4743 Sgr 2003.

Overall, the WT models with solar abundances represent the V4743 Sgr data better than current hydrostatic models with optimized abundances.
Though abudance optimizations will certainly further improve the fits, strong departures as determined with hydrostatic models are possibly not necessary.

With five parameters the WT models have more ``freedom'' than hydrostatic models, which have three.
The two extra parameters make the procedure for finding good fits slightly more complex, especially since their effects on the spectra partially overlap.
But we have presented a straight forward way to uniquely determine the best-fit model parameters, apart from the white-dwarf-mass-ambiguity in the WT models that leave $\Mwd$ undetermined.

Comparisons between models and SSS data should preferably be done in both ``as-observed'' and ``as-modeled'' representations, because each highlight different important aspects of the spectra.
Fits that only look good in one or the other representation are not acceptable.

The two properties most accurately constrained by the data are: 1) the ISM column depth, and 2) the bolometric luminosity (provided the distance to the source is known accurately).
The column depth can only be accurately determined for LETGS data, because the RGS wavelength coverage is insufficient.
We claim that precise determination of the ISM column depth is very important (or even crucial) if accurate fit parameters are to be obtained.
The additional freedom introduced by treating $\NH$ like a free parameter can easily double or triple the uncertainties in most model fit parameters.

\hspace{.5cm}
\acknowledgements
We thank Sumner Starrfield for the helpful discussions and comments on the draft of this manuscript, Jan-Uwe Ness for providing the observational data, and John Dombeck for providing his IDL color tables.

\bibliography{Nova, Phoenix}

\end{document}